\pdfoutput=1 
\documentclass[
  aps,%
  twocolumn,%
  prb,%
  groupedaddress,%
  superscriptaddress,%
  showpacs,%
  letterpaper,%
  amsfonts,%
  floatfix%
]{revtex4-1}

\RequirePackage{amsmath,amssymb,bm,wasysym}
\RequirePackage{pdfpages}
\RequirePackage{graphicx}
\RequirePackage{hyperref}
\makeatletter
\AtBeginDocument{\let\LS@rot\@undefined}
\makeatother
\hypersetup{%
  breaklinks = {true},
  citecolor = {black},
  colorlinks = {true},
  linkcolor = {black},
  pdfauthor = {\textcopyright\ Matthias Droth },
  pdfcreator = {\LaTeX\ and \flqq hyperref\frqq},
  pdffitwindow = {true},
  pdfmenubar = {true},
  pdfpagelayout = {OneColumn},
  pdfstartview = {FitH},
  pdftoolbar = {true},
  plainpages = {false},
  pdfpagemode = {UseThumbs},
}

\DeclareMathAlphabet{\mathpzc}{OT1}{pzc}{m}{it}
\newcommand{\Lietal}{Li \emph{et al.}}

\graphicspath{ {./Figs/} } 

\newcounter{topiccounter} 

\begin{document}
\title{Reply to ``Comment on `Piezoelectricity in planar boron 
nitride via a geometric phase'"}

\author{Matthias Droth}
\affiliation{Department of Physics, University of Konstanz, 
78457 Konstanz, Germany}
\affiliation{Laboratoire de Physique,
Ecole Normale Sup\'{e}rieure de Lyon, 69007 Lyon, France}

\author{Guido Burkard}
\affiliation{Department of Physics, University of Konstanz, 
78457 Konstanz, Germany}

\author{Vitor M.~Pereira}
\affiliation{%
  Department of Physics, National University of Singapore,
  2 Science Drive 3, Singapore 117542}
\affiliation{%
  Centre for Advanced 2D Materials, National University of Singapore,
  6 Science Drive 2, Singapore 117546}



\maketitle

\thispagestyle{plain}

In relation to our original paper [M.~Droth \emph{et al.}, Phys.~Rev.~B \textbf{94}, 
075404 (2016)], the Comment by \Lietal{} [Phys.~Rev.~B \textbf{98}, 167403 (2018)] 
claims to have identified a ``\emph{mistake in constructing the 
adiabatic process of the piezoelectricity}''. More specifically, they write that 
in our original work \emph{``the erroneous usage of the polarization difference 
formula in Eq. (4) leads to an invalid analytical expression of piezoelectric 
constant in Eq. (12) of Reference \onlinecite{Droth2016}''}. We explain below 
why these and other minor claims in the Comment are unwarranted, and why we 
maintain that our result is correct and physically sound.

\section{Choice of adiabatic parameter}%
The procedure used to obtain the piezoelectric constant in our original paper 
\cite{Droth2016} 
hinges on a calculation of the bulk polarization as an explicit function of the 
strain tensor [cf. the original Eq. (9)]. As established by the seminal work of 
Resta, Vanderbilt and King-Smith \cite{Resta,KingVanderbilt}, the only 
well defined physical quantity that is amenable to a controlled calculation is a 
variation in bulk polarization. This led to the development of the modern theory 
of polarization according to which such polarization differences are calculated 
by integrating the varying Berry curvature along an adiabatic path connecting 
the two states. One of the most typical situations involves computing the 
absolute bulk polarization of a given system: to exploit this method, one sets 
up an adiabatic path that connects that target with a 
reference system having zero bulk polarization \cite{Resta}. Such an adiabatic 
path is frequently defined in terms of a parameter of the Hamiltonian that can 
be continuously tuned between the initial and target state \cite{Resta} and, in 
particular, \emph{the adiabatic process does not have to necessarily reflect any 
real physical process}. For example, in first-principles implementations it is 
rather common to obtain the bulk polarization by an adiabatic process that 
continuously changes the atomic pseudopotentials, or which evolves the lattice 
from an inversion-symmetric 
configuration to the target state through a fictitious deformation path (for a 
specific example where this evolution of a ``virtual crystal'' is done for a 
BN-based system see Ref.~[\onlinecite{Nakhmanson}]). 

In calculations made with effective Hamiltonians, the earlier paper in the 
context of nanotubes by Mele and Kr\'{a}l \cite{Mele2002} relies on exactly the 
same approach that we used of evolving the gap parameter in an adiabatic path 
that begins with a strained carbon nanotube and ends with a strained BN 
nanotube. As our adiabatic parameter is the gap $\Delta$, our adiabatic process 
begins with strained graphene and ends with strained BN; strain is kept finite 
and constant throughout. The adiabatic process chosen by \Lietal{} \cite{Li2018} 
begins with 
relaxed BN and ends with strained BN. Since the bulk polarization of the initial 
state is zero in the two cases (in uniaxially strained graphene because of 
inversion symmetry, and in relaxed BN because of its $D_{3h}$ point group), they 
are, in principle, both legitimate adiabatic paths to determine the bulk 
polarization. It is therefore erroneous, and not in line with the modern 
formulation of the quantum theory of polarization, to claim that ``\emph{the 
correct adiabatic process of piezoelectricity should reflect the 
deformation-induced polarization difference from the initial state of undeformed 
h-BN to the final state of deformed h-BN}'', as stated in the Comment.

\section{The analytical expression for $e_{222}$}%
Reinstating the parameter $\kappa$ 
associated with the electron-phonon coupling \cite{Suzuura2002}, as per our 
original parametrization, \Lietal{}'s Eq.~(18) should read
\begin{equation}
e_{222}=\frac{e|\beta|\kappa}{2\pi a}\left[\text{sign}\,(\Delta)-
\frac{\Delta}{\sqrt{\sqrt{3} \pi t^2+\Delta^2}}\right].
\label{e222-Lietal}
\end{equation}
In the original paper, we obtain instead \cite{GapConvention} [cf.~our original
Eq.~(12)]
\begin{equation}
e_{222} = \frac{e|\beta|\kappa}{\pi^2 a}\tan^{-1}
\left[ \frac{\Delta}{\sqrt{2w^2+\Delta^2}} \right].
\label{e222-orig}
\end{equation}

Before further discussion, it is instructive to verify whether these results 
satisfy simple limits and symmetries. In the model Hamiltonian that is used to 
describe the BN monolayer \cite{Droth2016,Li2018}, changing $\Delta\to-\Delta$ 
amounts to an inversion transformation, since it is equivalent to swapping B 
with N atoms everywhere. Being a rank-3 tensor, $e_{222}$ should 
change sign under an inversion transformation, which is indeed satisfied by 
both results \eqref{e222-Lietal} and \eqref{e222-orig}.

However, it is clear that when $\Delta\to 0$, our result \eqref{e222-orig} 
decreases and ultimately becomes zero in the limit of graphene ($\Delta=0$). 
This is just as expected, because the six-fold rotational symmetry of the latter 
precludes a piezoelectric response ($e_{ijk} = 0$ in graphene by symmetry). In 
contrast, the expression \eqref{e222-Lietal} obtained by \Lietal{} remains 
finite as $\Delta\to 0$. In fact, according to their result, $e_{222}$ 
\emph{increases} monotonically when $\Delta$ decreases. In addition to failing 
to recover the graphene case in the limit $\Delta\to0$, 
this is contrary to physical intuition because, 
even if the gap remains finite, when it decreases, the charge transfer between B 
and N decreases as well. Even though in the modern theory the macroscopic dipole 
moment is not defined in terms of local properties, one expects its magnitude to 
qualitatively follow the trend of the bond polarity. Hence, the dipole moment 
induced by a finite deformation in this 
system is expected to decrease when the gap parameter $\Delta$ is brought to 
zero. Seeing this from another perspective, if one makes an infinitesimal 
perturbation to graphene that breaks its intrinsic sublattice symmetry, one is 
breaking the inversion symmetry of the system only by an infinitesimal amount as 
well, and expects the induced polarization to be small. But the result 
by \Lietal{} predicts otherwise, namely a gap-independent value of 
$\bm{P}$ (and, by extension, of $e_{222}$ as well) in the limit where the gap 
is much smaller than the bandwidth ($\Delta \ll t$), and a finite discontinuity 
in $e_{222}$ when $\Delta$ varies infinitesimally between $0^-$ and $0^+$.

Therefore, our original work is suitable to characterize, in an entirely 
analytical way, the behavior of both the electronic contribution to the bulk 
polarization and the piezoelectric constant in BN, as intended and stated in our 
original paper. 
Moreover, relying on an analytical model Hamiltonian, it is only useful if it 
captures the qualitative trends of these quantities, in particular when 
transitioning from BN to the graphene limit, which is not the case 
with the result in the Comment.

\section{Two different analytical results}%
There are two aspects related to the different results obtained for $e_{222}$ 
in Eqs.~\eqref{e222-Lietal} and \eqref{e222-orig}. The first one is a 
superficial difference related to the approximation used for the effective 
Brillouin zone (BZ) 
in the momentum integrations. Whereas our original paper uses a square, 
total-state-conserving effective BZ, the Comment uses a circular one. If 
$e_{222}$ is computed in a circular BZ domain (using polar coordinates for the 
integrations) we obtain
\begin{equation}
  e_{222} = \frac{e|\beta|\kappa}{4\pi a} \frac{\Delta}{\sqrt{q_c^2+\Delta^2}}
  .
\end{equation}
As expected, this expression agrees with Eq.~\eqref{e222-orig} to leading order 
when $\Delta \gg \{q_c,w\}$, up to a factor $\sim 1$ arising from the 
circular-vs-square geometry difference [note that $4w^2 = \pi q_c^2 = 
4\pi^2/(3\sqrt{3} a^2)$, see the original paper]. Moreover, it corresponds 
precisely \cite{GapConvention} to the second term in the result 
\eqref{e222-Lietal} above, which is the proposed analytical form in the Comment.

The second aspect has to do with the essential difference that makes the result 
proposed in the Comment a constant as $\Delta\to 0$, thus failing to capture 
the correct graphene (gapless) limit. The method followed in the Comment relies 
on two steps that are mathematically unsafe when combined: using strain as 
the adiabatic parameter and, in practice, differentiating the integral that yields 
the polarization with respect to strain beforehand. As a result, in the Comment, 
$e_{222}$ is expressed as the integral stated in their Eq.~(9), where the 
derivatives with respect to the strain components appear in the integrand. The 
problem with this approach is that the original integrand that yields the 
polarization $\bm{P}$ is not a continuous function over the domain of the 
multiple integral. Hence, ``passing'' the derivative from the outside to the 
inside of the multiple integral (Leibniz's rule) is not warranted. In Appendix 
\ref{app:integrations}, we show explicitly that, doing so, we obtain the result 
stated in the Comment (still using the gap as adiabatic parameter), but it 
arises only because of a blind application of Leibniz's rule.

In addition, to corroborate the mathematical robustness of our original 
result, we provide in the Supplemental Material \cite{Supplement} a 
\emph{Mathematica} notebook that verifies the steps involved in 
the calculations reported in our original paper.

\section{Symmetry constraints on $e_{ijk}$}%
The discussion presented in the Comment in connection to its Eqs.~(19-20) 
implicitly suggests that Eq.~(10) in our paper, which states 
\begin{equation}
  e_{211} = e_{112} = e_{121} = - e_{222},
  \label{symm-constraint}
\end{equation}
is incorrect and should instead read as the Comment's Eq.~(20). However, the 
relations \eqref{symm-constraint} \emph{must} be satisfied by \emph{any} rank-3 
physical tensor in a 
system containing a three-fold rotational axis, and with the mirror plane 
parallel to the direction $\mathbf{u}_2$ (refer to any textbook covering 
symmetry aspects of crystals, such as our original Ref.~[47]). This is the 
case of monolayer BN in the orientation shown in our Fig.~1, which has point 
group symmetry $D_{3h}$. Our definition in Eq.~(9) of the paper ensures 
$e_{ijk}$ is a Cartesian tensor.
Hence, it is clear that Eq.~(20) in the Comment violates the symmetry-imposed 
constraint among the nonzero components of $e_{ijk}$. The error stems from the 
flawed considerations in the text preceding Eq.~(19) of the Comment, namely 
because $\varepsilon_{ij}$ defined there is not a physical tensor (i.e., its 
components do not transform as those of a Cartesian tensor) due to the factor 
of 2 included in the definition of $\varepsilon_{12}$. 
Eq.~\eqref{symm-constraint} above is a relation involving components of a 
\emph{tensor} which should be adapted if one is using a Voigt representation, 
and not the other way around as stated in the Comment.

\section{Better agreement with the DFT result}
The authors of the Comment emphasize that they obtain a very good numerical 
agreement between their numerical result for $e_{222}$ and the one arising
from first-principles in a clamped-ion calculation of the electronic 
contributions associated with the $\pi$ and $\sigma$ bands \cite{Duerloo2012}. 
This is misleading because such an agreement is simply a numerical coincidence, 
especially in view of the approximations involved in the effective (Dirac) 
Hamiltonian and the uncertainty associated with the estimates of the logarithmic 
derivative of the hopping parameter under strain. Moreover, those 
estimates were not based on the specific bandstructure predicted by the quoted 
density functional theory (DFT) calculation.

The authors of the Comment also refer to Ref. [\onlinecite{newref}] as an 
alternative approach to check their calculation. However, piezoelectricity in 
BN is not discussed in that reference and it also does not offer an alternative 
approach. Instead, the calculation shown there is very similar to the one in 
their Comment, adapted to transition metal dichalcogenides.

\section{Conclusion}

In summary, we trust that our methodology and calculations reported 
originally in Ref.~[\onlinecite{Droth2016}] are sound and physically well 
justified.

MD acknowledges funding from the Deutsche Forschungsgemeinschaft (DFG, German 
Research Foundation) within Project No.~\mbox{317796071}. MD and GB thank the 
European Science Foundation and the DFG for support within the EuroGRAPHENE 
project CONGRAN and the DFG for funding within SFB 767 and FOR 912.

\appendix

\section{Differentiating beforehand}
\label{app:integrations}

There is an apparent advantage in performing the derivatives 
\begin{equation}
  \frac{\partial P_x}{\partial A_y}\biggr|_{\bm{A}=0} 
  \quad\text{and}\quad
  \frac{\partial P_y}{\partial A_x}\biggr|_{\bm{A}=0}\,,
\end{equation}
required to compute $e_{222}$, before performing the adiabatic and BZ integrals 
since, in this way, the integrand becomes a simpler algebraic function and the 
shifted Dirac point disappears. This is, in effect, what is implied by the 
procedure followed in the Comment by \Lietal{} when using strain as the 
adiabatic parameter.

However, there is a technical subtlety in doing so which we 
illustrate following our approach \cite{Droth2016} that uses the gap as the 
adiabatic parameter. The $i$-th Cartesian component of the polarization is 
obtained as
\begin{equation}
  P_i = \int_0^\Delta d\lambda \left[ 2e \sum_\tau \int_{\text{BZ/2}} 
  \frac{d\textbf{q}}{(2\pi)^2}\,\Omega_{q_i,\lambda}^{(\tau)} \right],
  \label{P-def-orig}
\end{equation}
which is our original Eq.~(4). To obtain the piezoelectric constant we need to 
compute, for example, [cf.~Eq.~(11) in the paper]
\begin{equation}
  e_{222} \propto \frac{\partial}{\partial A_x} 
    \left[
    \int_0^\Delta d\lambda \int_{\text{BZ}/2} \!\! d\bm{q} \,
    \Omega^{(\tau)}_{q_y,\lambda}
    \right]_{\bm{A}=0}.
  \label{e222-dA-outside}
\end{equation}
Whether or not one can safely employ Leibniz's rule and pass the derivative 
to the inside of the integral depends on $\Omega^{(\tau)}_{q_y,\lambda}$ 
being a continuous function of $(q_x,q_y,\lambda)$ in the domain of the 
multiple integral. However, 
\begin{equation}
  \Omega^{(\tau)}_{q_y,\lambda} = \frac{\tau}{2} \,
  \frac{q_x-A_x}{[(q_x-A_x)^2+(q_y-A_y)^2+\lambda^2]^{3/2}}\label{BC_app}
\end{equation}
has a clear discontinuity at the point $(q_x,q_y,\lambda)=(A_x,A_y,0)$ and, 
therefore, the conditions of Leibniz's rule are not fulfilled [the singularity is 
integrable though\,---\,to zero\,---, as we demonstrate in Appendix 
\ref{app:singularity}]. 
As a result of this discontinuity, even though the integral of $\partial 
\Omega^{(\tau)}_{q_y,\lambda}/\partial A_x$ converges, \emph{its value depends 
on the sequence of the partial integrations}. We now show this explicitly.

We are interested in the integral
\begin{equation}
  I \equiv \int_0^\Delta d\lambda \int_{\text{BZ}/2} d\bm{q} \,
  \frac{\partial \Omega^{(\tau)}_{q_y,\lambda}}{\partial A_x} 
  \Biggr|_{\bm{A}=0},
  \label{I-def}
\end{equation}
where
\begin{equation}
  \frac{\partial \Omega^{(\tau)}_{q_y,\lambda}}{\partial A_x} 
  \biggr|_{\bm{A}=0} =
  \frac{2q_x^2-q_y^2-\lambda^2}{(q_x^2+q_y^2+\lambda^2)^{5/2}}
  .
\end{equation}
Since we are focusing on the technical aspect here, for simplicity, we 
drop all the non-essential prefactors and, for definiteness, consider 
$\Delta>0$ in the remainder of this appendix. 
Computing \eqref{I-def} by integrating first over the BZ and $\lambda$ 
last, we obtain 
\begin{subequations}
\begin{equation}
  I_{\lambda\text{ last, }\square} = -2\, 
  \text{tan}^{-1} \!\! \left( \frac{\Delta}{ \sqrt{2w^2+\Delta^2} } \right).
  \label{e222-cart}
\end{equation}
With the prefactors, this is our result (12) in the original paper, where the BZ 
integral has been performed over a square-shaped, state-conserving 
BZ, as discussed in the paper. For completeness, if one opts to use a circularly 
shaped domain with cutoff momentum $q_c$ (the choice made in the Comment), the 
result for $I$ becomes
\begin{equation}
  I_{\lambda\text{ last, }\bigcirc} = 
    - \frac{\pi\Delta}{ 2\sqrt{q_c^2+\Delta^2} } .
  \label{e222-polar}
\end{equation}
\end{subequations}

On the other hand, if one performs the \emph{adiabatic integral first} followed 
by that over the BZ, one obtains 
\begin{subequations} \label{lambda-first}
\begin{equation}
  I_{\lambda\text{ first, }\square} = 
  \pi - 2\,\text{tan}^{-1} \left( \frac{\Delta}{ \sqrt{2w^2+\Delta^2} } \right)
  ,
  \label{e222-cart-2}
\end{equation}
for the square BZ, and
\begin{equation}
  I_{\lambda\text{ last, }\bigcirc} = 
    \frac{\pi}{2} - \frac{\pi\Delta}{ 2\sqrt{q_c^2+\Delta^2} }
  \label{e222-polar-lambda}
\end{equation}
\end{subequations}
for the circular BZ. Note that this last result is the one reported in 
Eq.~(18) of the Comment once the prefactors are restored 
\cite{GapConvention}.
Therefore, computing \eqref{I-def} by performing the adiabatic integral 
\emph{first} followed by that over the BZ, introduces a constant, independent of 
$\Delta$. 

Mathematically, this dependence of the result on the order of the 
multiple integrations arises because Eq.~\eqref{I-def} does not converge 
absolutely and, consequently, it is sensitive to whether one performs the $q_x$ 
integration before or after the one over $\lambda$ (changing to polar 
coordinates does not change this, of course) \cite{Pathology}.
This non-uniform convergence is, of course, a consequence of having 
erroneously used Leibniz's rule when the integrand of \eqref{e222-dA-outside} 
is not a continuous function (incidentally, performing the $\lambda$ integral 
last, which is physically motivated, yields the correct result).

In order to avoid these pitfalls, the method of calculation used in our 
original paper keeps $A_x$ and $A_y$ finite until the triple integral that 
yields $\bm{P}$ is obtained. We only linearize the result in $\bm{A}$ 
afterwards. In this way, the result is mathematically well defined, 
irrespective of the sequence used for the triple integration \cite{Supplement}. 


\section{Singularity of the Berry curvature}
\label{app:singularity}

In spherical coordinates,
\begin{eqnarray}
q_x{-}A_x&=&r\sin(\theta)\cos(\phi)\,,\nonumber\\
q_y{-}A_y&=&r\sin(\theta)\sin(\phi)\,,\\
\lambda&=&r\cos(\theta)\,,\nonumber
\end{eqnarray}
the Berry curvature given in Eq.~(\ref{BC_app}) reads
\begin{equation}
\Omega_{q_y,\lambda}^{(\tau)}\propto
\frac{r\sin(\theta)\cos(\phi)}{r^3}=
\frac{\sin(\theta)\cos(\phi)}{r^2}\,.
\end{equation}
This expression has a discontinuity at $r{=}0$, i.e., at 
$(q_x,\,q_y,\,\lambda){=} (A_x,\,A_y,\,0)$, which lies in the 
integration domain of Eq.~(\ref{P-def-orig}). The 
support of this point is $0$ but the Berry curvature diverges so one must, in 
principle, verify whether there is a finite contribution to the integral. This 
contribution can be checked as follows:
\begin{eqnarray}
&&\lim_{{\delta}\to0}\int_0^{\delta}{\rm d}r\int_0^{\pi/2}
{\rm d}\theta\,r\int_0^{2\pi}{\rm d}\phi\,r\,\sin(\theta)\,
\Omega_{q_y,\lambda}^{(\tau)}\nonumber\\
&&\propto
\lim_{\delta\to0}\int_0^\delta\frac{r^2}{r^2}{\rm d}r
\cdot\int_0^{\pi/2}\sin^2(\theta)\,{\rm d}\theta\cdot\int_0^{2\pi}
\cos(\phi)\,{\rm d}\phi\nonumber\\
&&=\lim_{\delta\to0}(\delta-0)\cdot\frac{\pi}{4}\cdot0\nonumber\\
&&=0\,.
\end{eqnarray}
This means that the singularity at the shifted Dirac points integrates to 0 
and does not add any particular contribution to the result of 
Eq.~(\ref{P-def-orig}).


\newpage\includepdf{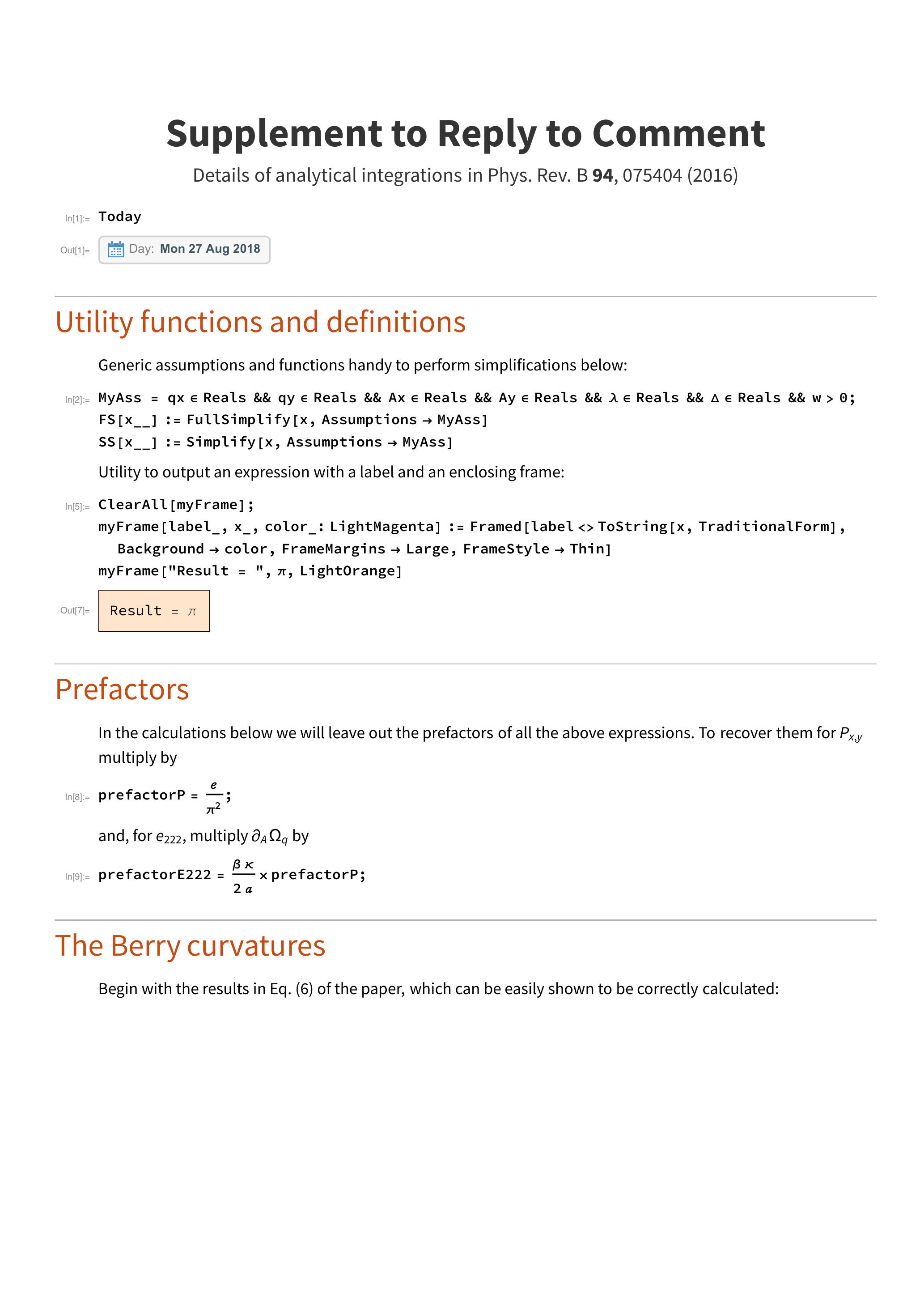}$\text{}$
\newpage\includepdf{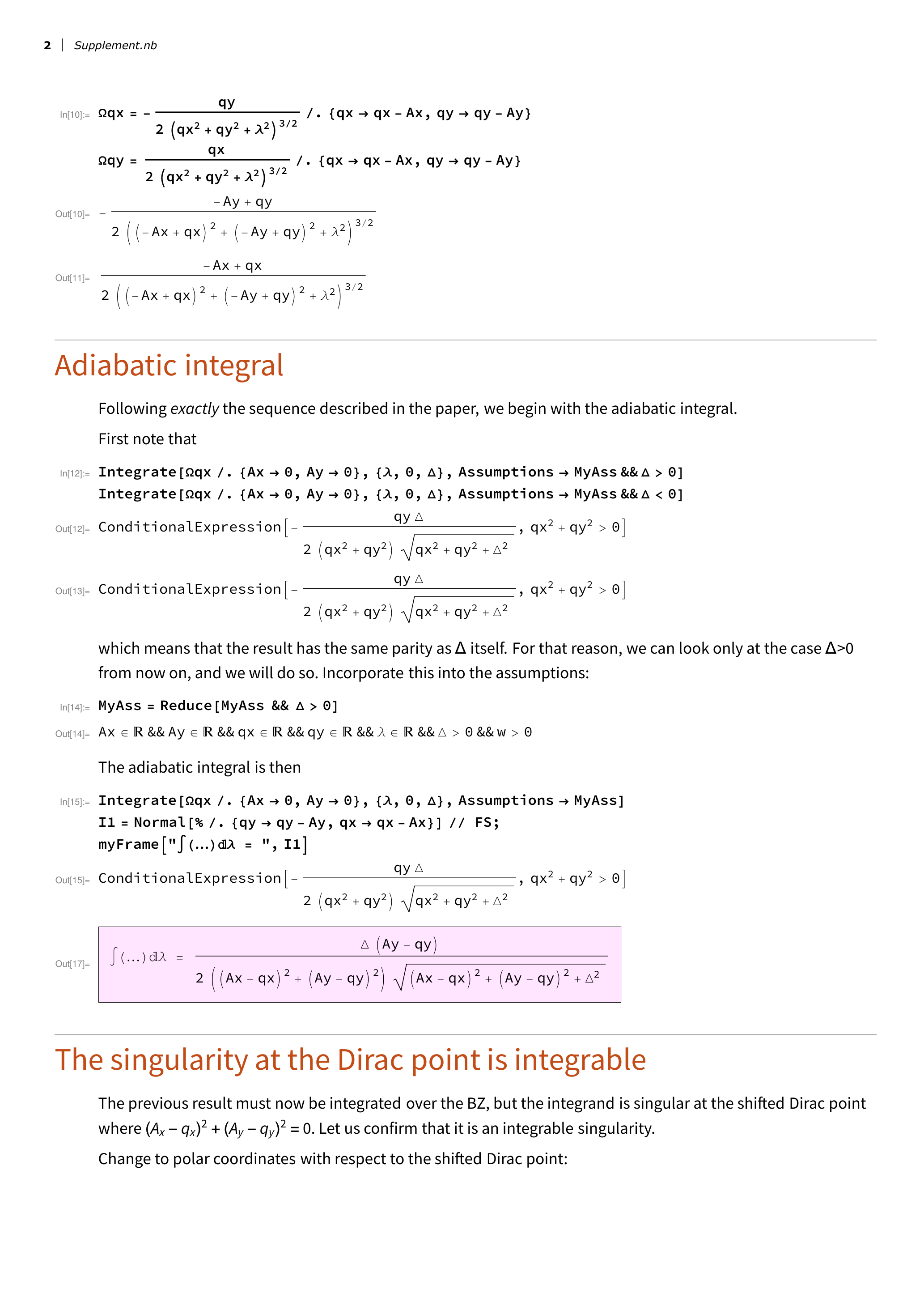}$\text{}$
\newpage\includepdf{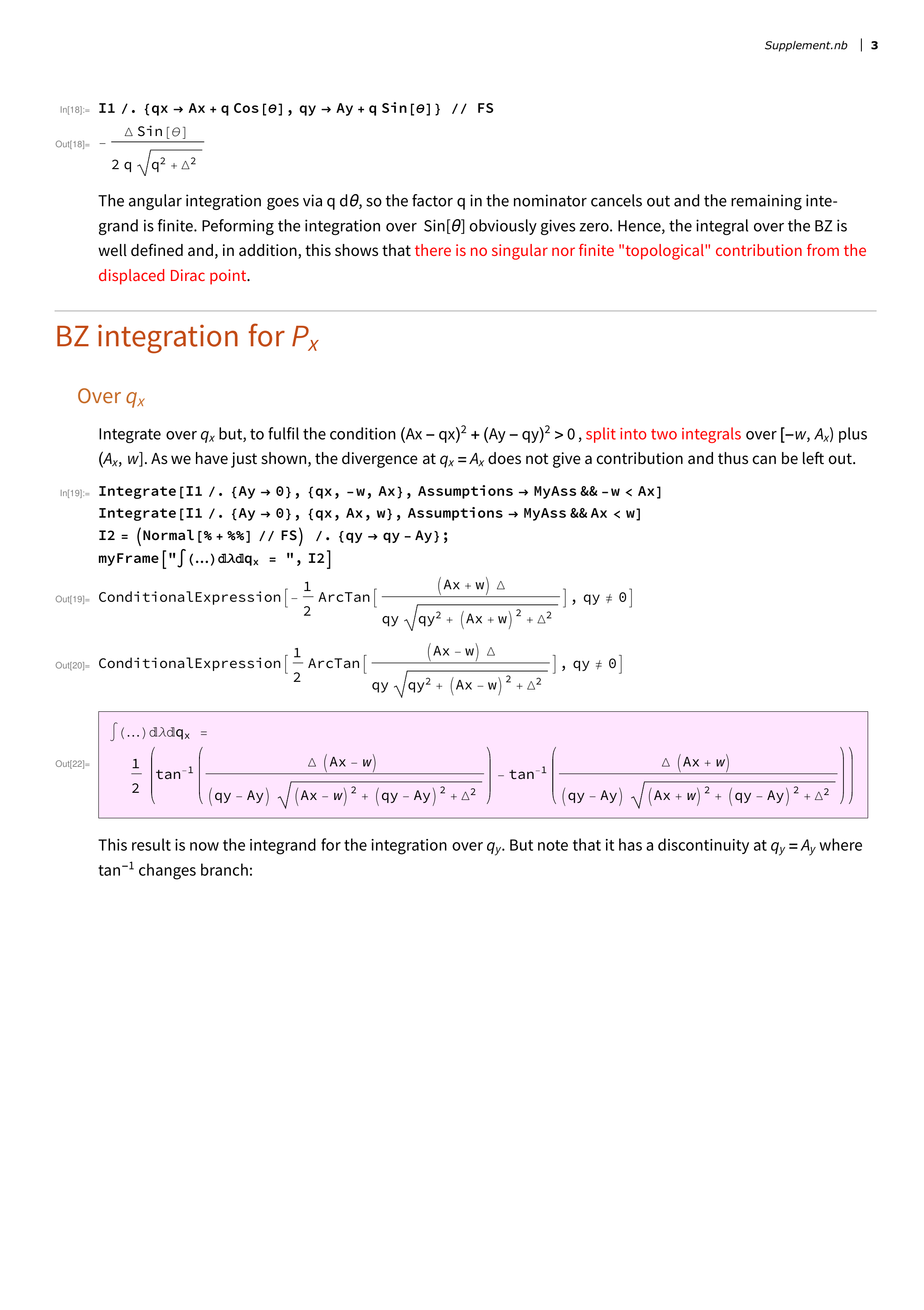}$\text{}$
\newpage\includepdf{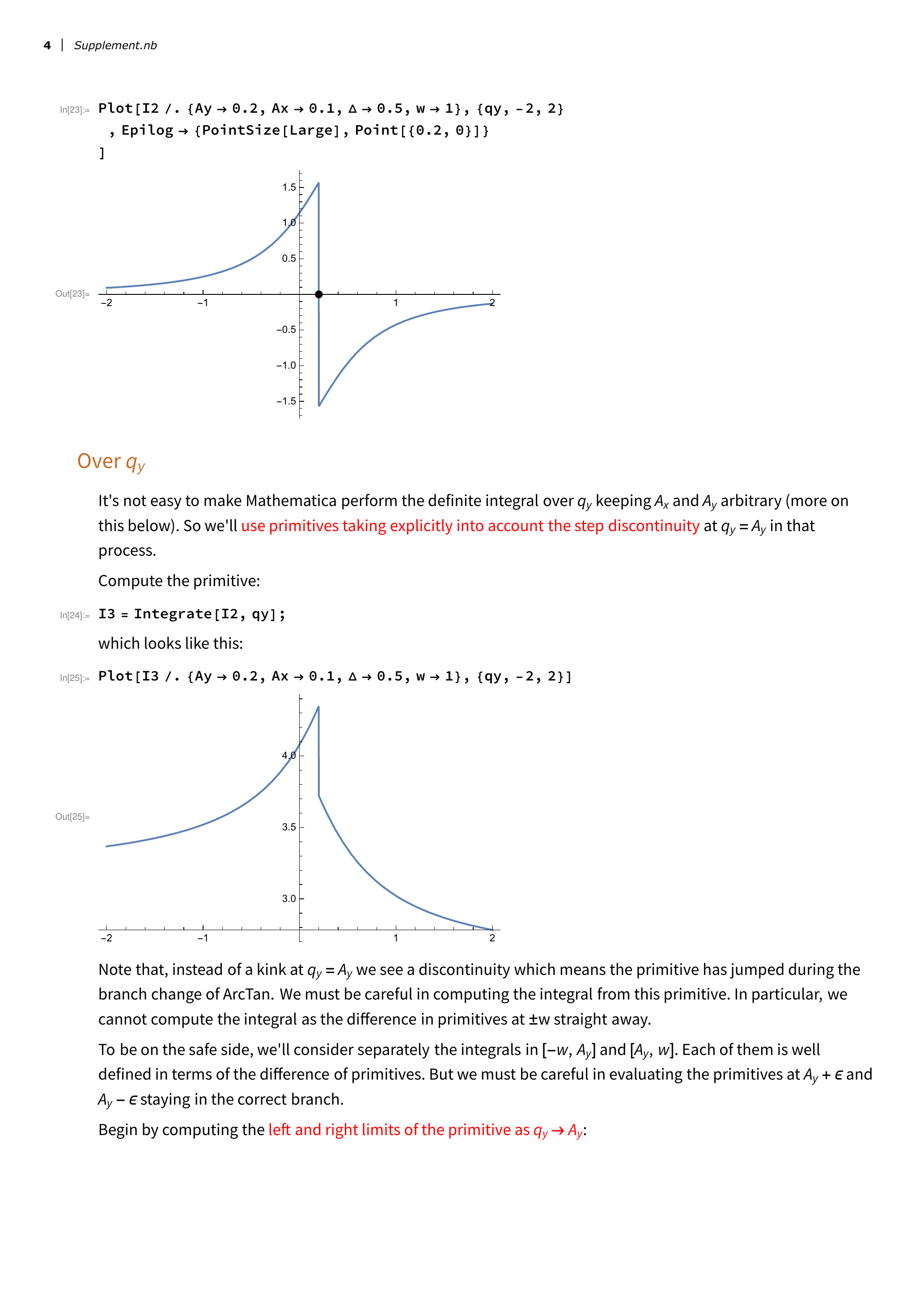}$\text{}$
\newpage\includepdf{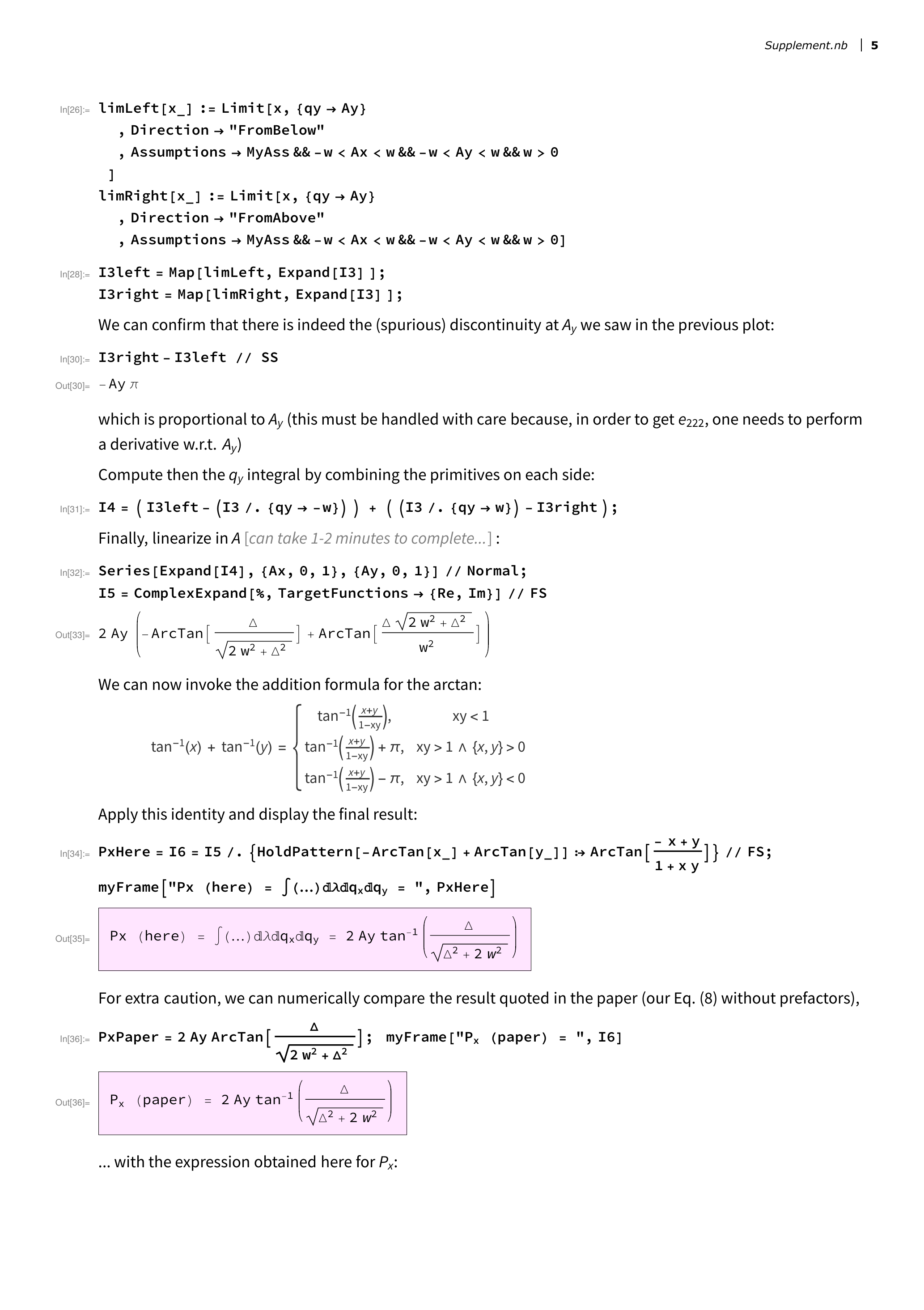}$\text{}$
\newpage\includepdf{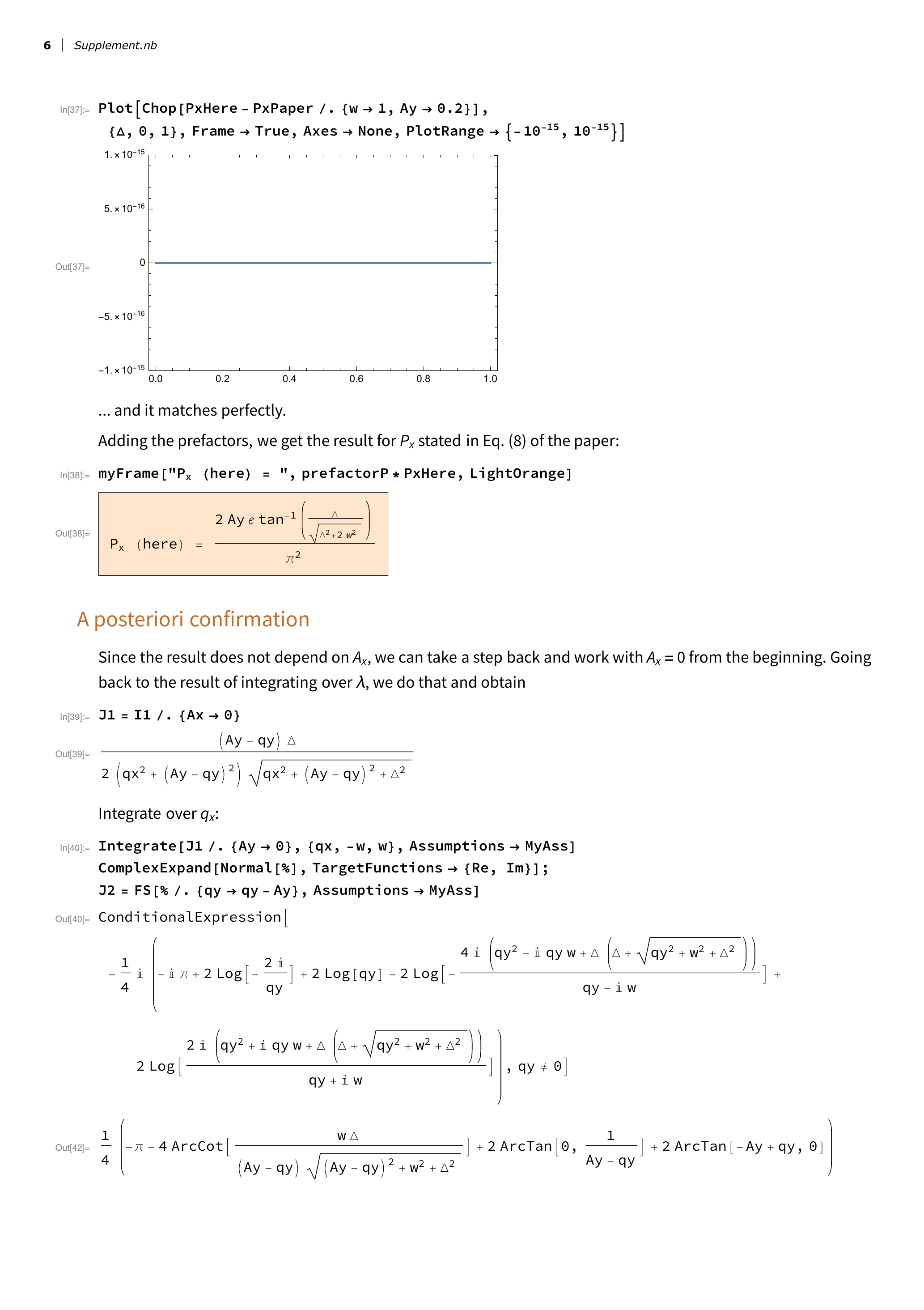}$\text{}$
\newpage\includepdf{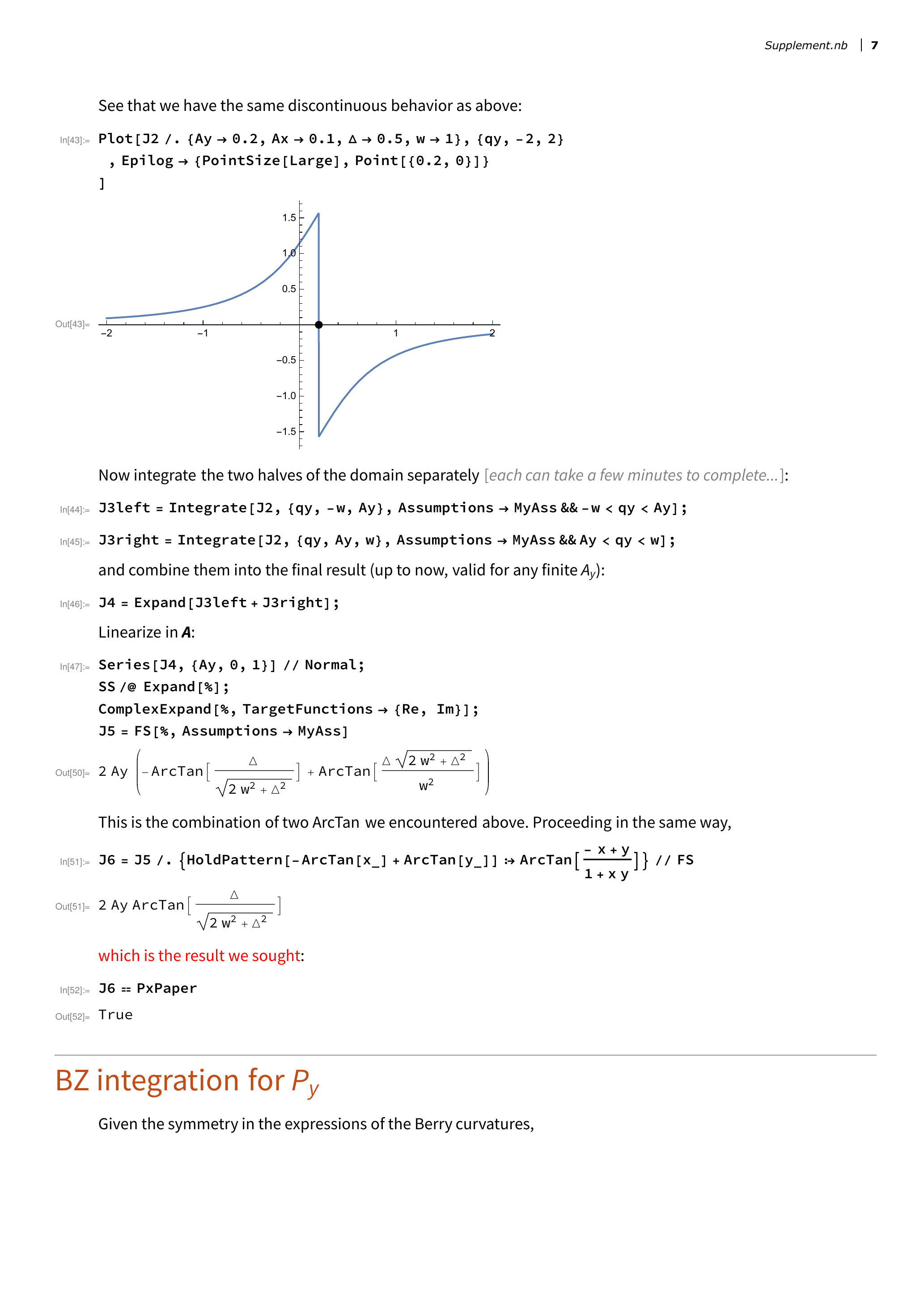}$\text{}$
\newpage\includepdf{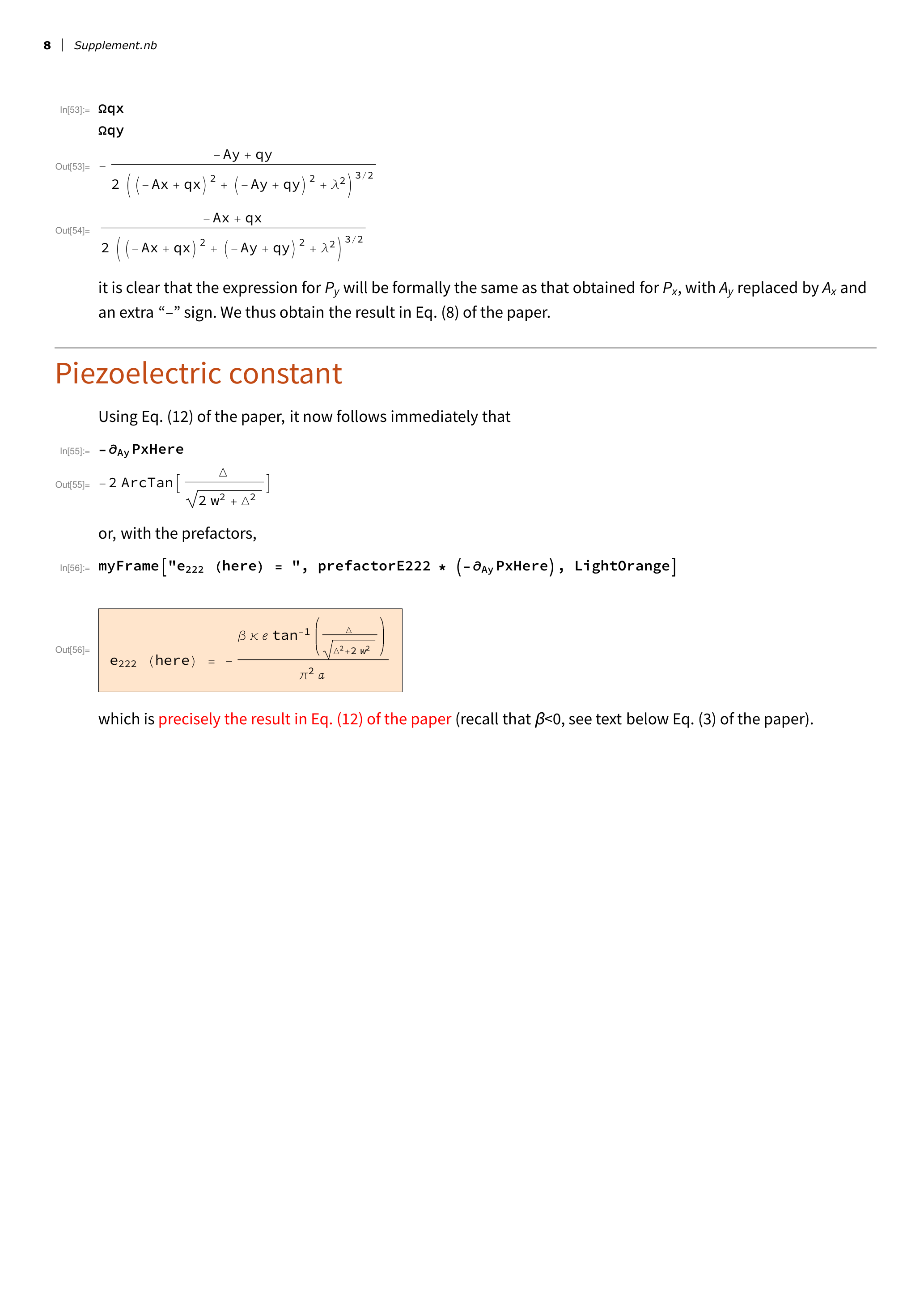}$\text{}$

\begin{thebibliography}{}
\bibitem{GapConvention}
When comparing, note that in our notation the gap is $\hbar v_F \Delta$, while 
it is simply represented by $\Delta$ in the notation of the Comment (we 
maintain the original notations of each).
\bibitem{Pathology}
The integral in question has precisely the same pathology as the 
double integral $\int_0^1\int_0^1 \frac{x^2-y^2}{(x^2+y^2)^2}\,dx\,dy$, which 
is a common counterexample to the interchangeability in the order of 
integrations in the absence of absolute convergence. In this case, the result 
is $+\pi/4$ ($-\pi/4$) if $y$ ($x$) is integrated first.
\end{thebibliography}
\end{document}